# Negative Spin Valve effects in manganite/organic based devices

A. Riminucci, I. Bergenti, L. E. Hueso, M. Murgia, C. Taliani, Y. Zhan, F. Casoli[a],

M. P. de Jong[b], V. Dediu

*ISMN-CNR, via Gobetti 101, 40129 Bologna, Italy*

*[a]IMEM-CNR, Parco Area delle Scienze 37/A, 43100 Parma, Italy*

*[b]Department of Physics, IFM, Linköping University, S-581 83 Linköping, Sweden*

We report detailed investigations of hybrid organic-inorganic vertical spin valves. Spin polarized injection in tris(8-hydroxyquinoline) aluminum (Alq$_3$) organic semiconductor (OS) was performed using La$_{0.7}$Sr$_{0.3}$MnO$_3$ manganite as the bottom electrode and Co as the top electrode. While manganite was directly connected to the organic semiconductor layer, a thin tunnel barrier was placed between the OS and the Co electrode. A clear negative spin valve effect – low resistance for antiparallel electrodes configuration – was observed below 210 K in various devices using two different tunnel barriers: LiF and Al$_2$O$_3$. The magnetoresistance effect was found to be strongly asymmetric with respect to the bias voltage. Photoelectron Spectroscopy (PES) investigation of the interface between manganite and Alq$_3$ revealed a strong interface dipole, which leads to a better matching of the metal Fermi level with Alq$_3$ LUMO (1.1 eV) rather than with HOMO level (1.7 eV). This unequivocally indicates that the current in these devices is dominated by the electron channel, and not by holes as previously suggested. The knowledge of the energy diagram at the bottom interface allowed us to work out a semi- quantitative model explaining both negative spin valve effect and strong voltage asymmetry. This model involves a sharp energy selection of the moving charges by the very narrow LUMO level of the organic material leading to peculiar resonant effects.

Spin dependent transport has been the object of intense research since the demonstration of magnetoresistive effects in magnetic tunnel junctions and metallic multilayers [1, 2]. The field has evolved to the extent of producing commercial applications for magnetic recording and memory. However, coherent spin transport on distances over the nanometre scale has proved so far quite difficult in normal metals and semiconductors [3]. This difficulty has motivated the search for new materials in which both efficient spin injection and transport can be achieved. Among others, $\pi$-conjugated organic semiconductors (OS) have emerged as suitable candidates, mainly thanks to their low spin-orbit interactions and their proved feasibility to be integrated in partially inorganic devices [4, 5].

Spin injection into organic semiconductors was first demonstrated in lateral devices with highly spin polarized manganite $La_{0.7}Sr_{0.3}MnO_3$ (LSMO) electrodes and sexythiophene (T6) as channel material [4]. Subsequently, magnetoresistance (MR) in vertical organic spin valves (OrgSVs) with LSMO and Cobalt electrodes was observed, using tris(8-hydroxyquinoline) aluminum ($Alq_3$) as interlayer [6, 7]. A recent communication reported room temperature spin polarized tunneling across $Alq_3$ layer [8]. To date, the combination of LSMO and Alq3 is one of the most successful in the organic spintronics field and definitely requires deeper qualitative and quantitative understanding.

One issue in need of further research is the energetics at the most crucial parts of the device – the interfaces –, for which information is totally lacking. The use of unperturbed energy level alignment at LSMO/$Alq_3$ interfaces is obviously not sufficient. It leads automatically to the conclusion of hole conductivity in $Alq_3$ based OrgSVs [9], thus contradicting the well-established n-type nature of this material [10]. Another important experimental result is the negative magnetoresistive effect (low resistance state for antiparallel configuration) which has been systematically detected in LSMO/Alq3/Co heterostructures, while its explanation is still controversial. A preliminary explanation involves the opposite spin polarization at the Fermi energy of the manganite and the d-band electrons in Co but does not take into account the role of s-electrons [9]. The latter have been repeatedly detected in various (mainly tunnel) devices [11, 12] and need therefore to be taken into account and compared to the d-channel.

In this letter we report on new results shedding light on some peculiar properties of organic spin valves. In particular, voltage dependent magnetoresistance and information coming from Photoelectron Spectroscopy (PES) characterizations of the LSMO/OS interface allow us to build up a more realistic energy diagram and bands/level alignment in these devices. This diagram, though still approximate (for the complete diagram detailed investigations of *all* the interfaces are required), gives us the opportunity to understand some peculiar properties of OrgSVs, such as the negative magnetoresistive effect and carrier nature.

Figure 1 shows a schematic diagram of the OrgSV investigated in this paper. The bottom electrode is an epitaxial manganite film, $La_{0.7}Sr_{0.3}MnO_3$ (LSMO), followed by an $Alq_3$ spin/charge transport layer, a thin tunnel barrier and a Co top electrode. LSMO films (20 nm thick) were grown by pulsed electron deposition (pulsed plasma enhanced configuration - PPD) on matching perovskite substrates ($NdGaO_3$ or $SrTiO_3$). PPD, also called channel spark ablation, has been extensively used for the growth of various oxide films [13, 14]. Detailed surface X-ray photoemission spectroscopy (XPS, XAFS) showed surface metallicity (detectable Fermi level) and strong circular magnetic dichroism up to room temperature [15]. Both XPS and UPS techniques were used for detecting the work function for stoichiometric surface state: values of 4.8-4.9 eV were identified [15, 16]. Coercive fields measured by Magneto-Optical Kerr Effect (MOKE) were typically in the range of 10-20 Oe at room temperature, while increasing up to 300 Oe at 60 K.

In this report we concentrate on tris(8-hydroxyquinoline) aluminum ($Alq_3$) OrgSVs. $Alq_3$ is a π-conjugated molecule with remarkable properties in organic luminescent devices like OLEDs. It consists of one aluminium ion ($Al^{3+}$) and three 8-hydroxyquinoline molecules in which one nitrogen atom and one oxygen ion for each ligand coordinates the $Al^{3+}$ ion with a pseudo-octahedral structure. The electronic structure of $Al^{3+}$ is $1s^2 2s^2 2p^6$, similar to that of an inert gas atom, leading to the $Alq_3$ high stability in dry atmosphere. Vacuum-sublimed thin films are typically amorphous and have been shown to be morphologically stable at room temperature [10, 17]. It is important to note that $Alq_3$ is an electron charge carrier, while most OS are hole conductors. The electron mobility in this material is almost two orders of magnitude higher than hole mobility [18].

In our devices 100 or 200 nm of $Alq_3$ were deposited by Organic Molecular Beam Deposition (OMBD) in UHV conditions ($10^{-9}$-$10^{-10}$ mBar) on LSMO thin layers. Prior to deposition the LSMO surface was reconstructed following the annealing procedures established by XPS investigations [15]. Between the OS and the Co layer we deposited a 2 nm thick $Al_2O_3$ or LiF tunnel barrier by PPD or OMBD respectively. The 35 nm thick top Co electrode was deposited by RF sputtering. The choice of $Al_2O_3$ was based on its well-known properties as a tunnel barrier in magnetic tunnel junctions, while LiF was studied mainly because of it successful use in OLEDs.

The OrgSVs were measured in magnetic fields up to 3 kOe and in a temperature range 40-300 K at various voltages. The magnetoresistance of each individual electrode was carefully studied, enabling us to rule out anisotropic MR as the origin of our findings.

Before analyzing magnetoresistance in our hybrid devices, detailed current (I)-voltage (V) characterizations were performed. In the description that follows, a positive voltage means that the LSMO electrode is positive relative to the Co electrode. Typical IV curves show (Fig. 2) strongly non linear and anisotropic behavior with an insulating-like temperature trend. This insulating behavior rules out the contribution of eventual shorts circuit channels in these devices [19]. The non linearity in Fig. 2 provides a first indication for the existence of strong interface barriers, whose nature will be discussed below, while the asymmetric shape is an obvious result of the device asymmetry.

In order to estimate the barrier height and width we performed a fitting of the IV curves. We used a tunneling model supposing a trapezoidal barrier of heights $\varphi_1$ and $\varphi_2$ and width $d$ [20]. Following this model, the conductance should follow a polynomial dependence:

$$\frac{G(V)}{G(0)} = 1 - \left(\frac{A_0 \Delta\varphi}{16 \bar{\varphi}^{3/2}}\right) eV + \left(\frac{9}{128} \frac{A_0^2}{\bar{\varphi}}\right)(eV)^2$$

where $\Delta\varphi = \varphi_2 - \varphi_1$, $A_0 = 4(2m)^{1/2} d / 3\hbar$, and $G(0) = (3.16 \times 10^{10} \bar{\varphi}^{1/2} / d) \exp(-1.025 d \bar{\varphi}^{1/2})$.

The conductances at positive and negative voltages were fitted independently, and only small differences in the fitting parameters could be extracted. A fit to a low temperature IV characteristic with $\bar{\varphi} = 1V$, $d = 30\,\text{Å}$ and $\Delta\varphi = 0.46V$ can be seen in Figure 2. From fits to curves corresponding to different devices measured at different temperatures we extracted very similar parameters, showing the robustness of the transport mechanisms. The low bias resistivity of our devices ($10^5$ Ohms.cm) is characteristic for electron transport in Alq$_3$ layers.

We measured the magnetoresistance of several LSMO/Alq$_3$/Tunnel Barrier/Co devices. The tunnel barrier was LiF in some devices and Al$_2$O$_3$ in others. We routinely detected a negative SV effect in such devices up to 210 K regardless of the kind of tunnel barrier between the organic layer and the Co electrode. A typical SV magnetoresistance at 100 K for the LSMO/Alq$_3$(200nm)/Al$_2$O$_3$/Co vertical device is shown in Fig. 3. At high field, the magnetizations of the two electrodes are parallel. As the field is lowered and eventually reversed, we reach the coercive field of the LSMO electrode, and the system switches to antiparallel configuration. As the field is increased further and becomes greater than the coercive field of the Co electrode, the system goes back to parallel configuration. The value of the coercive field of the Co film measured in this way (~1500 Oe at 60 K) is considerably higher than the value measured by Magneto-Optical Kerr Effect (MOKE). This may be due to the fact that the magnetic behavior of the Co/OS interface is different from that at the top surface, as reported for example in [21]. In addition, a kind of double switching at "high fields" can be observed, confirming the possible magnetic difference between bulk and interface cobalt regions.

The main output of these magnetoresistance studies is the inverse spin valve effect in devices containing tunnel barrier separation. It is important to note that this inverse sign was found independently on the choice of the tunnel barrier material.

As the focus of this paper is the study of interfaces, we have not yet performed a systematic investigation of the spin valve effects dependence on the OS thickness. Although the samples with 200 nm thick $Alq_3$ layer showed lower magnetoresistance, we believe that a systematic study of the thickness dependence would require additional efforts to achieve better control and definition of the top interface.

Figure 4 shows the dependence of the SV effect on the direction and magnitude of the current injected during the measurement. At negative voltages (i.e., the LSMO layer is negative with respect to the Co layer) the SV effect is greater than when the voltage is positive. The highest magnetoresistance value in Figure 4 corresponds to the case of electron/spin injection from the manganite towards the Co electrode, while a smaller SV effect is observed when the current is reversed.

The negative SV effect and the strong bias asymmetry seem to be particularly interesting. As mentioned above, a negative SV effect has been reported on similar organic SVs [6, 7], and the explanation was based on the negative (spin down) polarization of the d-electrons in Co. Although qualitatively correct, this first explanation did not take into account the possible effect of the s-band, which is positively spin polarized and was found to play a dominant role in various Co based tunneling devices [22]. Special selection rules should operate in the device in order to provide higher efficiency of the d-electrons coming from the transition metal electrode.

The negative spin polarized injection form Co surface was also observed in some inorganic devices like LSMO/SrTiO₃/Co tunnel junctions [12]. In that case it was explained through an unusual interface magnetic configuration between Co and $SrTiO_3$ [22, 23]. Indeed, the spin valve effect switched into positive after the substitution of the $SrTiO_3$ barrier with $Al_2O_3$ [12]. By contrast, in our case the negative SV effect was observed for two different tunnel barrier materials, LiF and $Al_2O_3$, and even with no tunnel barrier at all [6].

In order to shed light on these effects we performed a detailed PES analysis of the LSMO/$Alq_3$ interface, one of the basic elements of our devices [16]. PES investigations revealed the presence of a strong interface dipole of about 0.9 eV, which shifts considerably the energy levels of the system.

Figure 5 shows the distribution of the density of states and the energy level alignment across the device starting from the manganite electrode placed on the left hand side. The figure displays both values directly measured in our experiments and values estimated from reasonable approximations whose validity is discussed below. The values directly measured by the PES technique (highlighted by circles) are: the LSMO workfunction of 4.9 eV,  the $E_F(LSMO) -$

HOMO(Alq$_3$) energy difference of 1.7 eV, and the dipole energy of 0.9 eV. In addition to these data the LUMO(Alq$_3$) value is calculated from the optical gap (2.8 eV), and Fermi level alignment with Co DOS across the tunnel barrier is assumed. The latter assumption seems quite reasonable, as no strong dipole is expected when an insulator is placed between a metal and an organic semiconductor [24].

Although the use of the optical gap to estimate the transport gap in organic semiconductors, i.e. the HOMO-LUMO splitting, has been widely applied (see for example the review by Ishii et al. [25]), it gives an underestimated value due to the excitonic nature of the optically excited states. The transport gap is thus larger than the optical gap by the exciton binding energy. Therefore, the position of the LUMO level as shown in Fig. 5 is at the lowest possible value above the HOMO. Nevertheless, the barrier height of about 1 eV (very close to the optical gap approach) for the injection into the LUMO level is strongly supported by the transport characterizations as well as by the resistivity values of our devices ($10^5$ Ohms cm), indicating electron rather than hole transport.

We can see that the presence of the dipole at LSMO/Alq3 interface has a strong impact on the injection from the manganite electrode: the barrier between LSMO $E_F$ and the LUMO level (1.1 eV) is now much lower than $E_F$ − HOMO barrier (1.7 eV). While the "non-distorted" picture [6, 7] suggests the injection into HOMO level, the real situation at this interface clearly indicates the electron injection and electron transport: the barrier difference strongly favors the injection into LUMO states.

The energy diagram in figure 5 provides a key to understand the origin of the negative sign of the SV effect for the electrons flowing from LSMO towards Co. We do not need any additional information about LSMO except its workfunction and the positive sign of its spin polarization, which is universally accepted. The Co density of states is taken from the review of Tsymbal et al. [22] and represents the bulk d-states for Co hcp configuration. The most striking feature of the whole energy diagram is the presence of a strong resonance between the LUMO level of Alq3 and the sharp "spin down" peak of Co at $E_F$+1eV. Although further detailed characterizations of the second interface are required for a more accurate definition of the absolute values estimated in Fig. 5, our tentative diagram points towards the need to take into account the resonant connection between the LUMO level and the sharp peaks of the Co d-band. In the framework of the Fig. 5 approximation, the resonance observed gives a satisfactory explanation for the negative SV effect through the increased strength of the d-channel over the s-electrons band. At the present state of the art it is not possible to describe in detail the current/spin flow in the opposite direction, but the diagram in Fig. 5 clearly indicates a strong device asymmetry in agreement with Fig. 3.

The charge and the spin injection in our OrgSVs are radically different from those in conventional inorganic tunnel junctions and spin injection devices. In the OrgSV the two spin polarized reservoirs are connected by a very narrow (hopping) channel at LUMO (HOMO in some other OS) energy. The LUMO channel is not represented by a real conducting band, but rather by a pseudo-localized level of 0.1 eV width [26]. This channel performs a strong energy selection as no injection is possible into the OS gap, at least in elastic approximation. Thus, in our spin valves we have a first tunneling event from the LSMO into the localized state, followed by hopping conductivity across the "thick" (100 nm) OS layer and, subsequently, a second tunneling effect from the localized state in the OS across the second barrier into the Co states. To our best knowledge, no theories describing the spin polarization for these cases are yet available.

To conclude, we detected an inverse spin valve effect in LSMO/Alq$_3$/Co vertical devices with an artificial tunnel barrier placed at top Alq$_3$/Co interface. Two different materials used for tunnel barriers, Al$_2$O$_3$ and LiF, showed similar results with no qualitative differences, while the strength of the magnetoresistance effect was found to be stronger for the Al$_2$O$_3$ barrier. Furthermore, strong voltage asymmetry was found, as the magnetoresistance is significantly stronger for electrons (spins) injected from the LSMO towards Co. The PES characterization of the LSMO/Alq$_3$ interface was used for the development of a semi-quantitative model which gives a good explanation for both the inverse spin valve effect and the strong device asymmetry.

**CITATIONS**

**Figure Captions**

*Figure 1*. Schematic plot of a vertical organic spin valve. The bottom electrode is the ferromagnetic manganite LSMO, while the top one is Cobalt. In between the electrodes is located the organic semiconductor (OS), while an artificial tunnel barrier is introduced between this and the top electrode.

*Figure 2*. (Main) Current (I) –Voltage (V) characteristic of a typical organic spin valve. Both the voltage asymmetry and the insulating behaviour with temperature can be observed from the curves. (Inset) Normalized conductance at 100K (circles), together with the fit to the Brinkman-Dynes-Rowell model (line).

*Figure 3*. Low temperature magnetoresistance versus magnetic field for a typical LSMO/Alq$_3$/Al$_2$O$_3$/Co organic spin valve

*Figure 4*. Variation of the magnetoresistance of an organic spin valve with the applied voltage. The asymmetric response is clearly observed.

*Figure 5*. Proposed energy diagram for a LSMO/Alq$_3$/Tunnel Barrier/Co organic spin valve

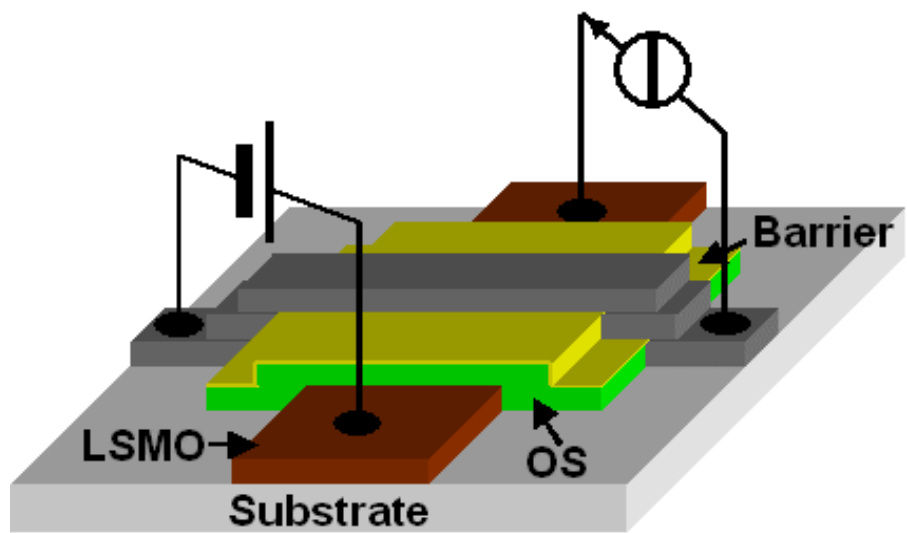

*Figure 1*

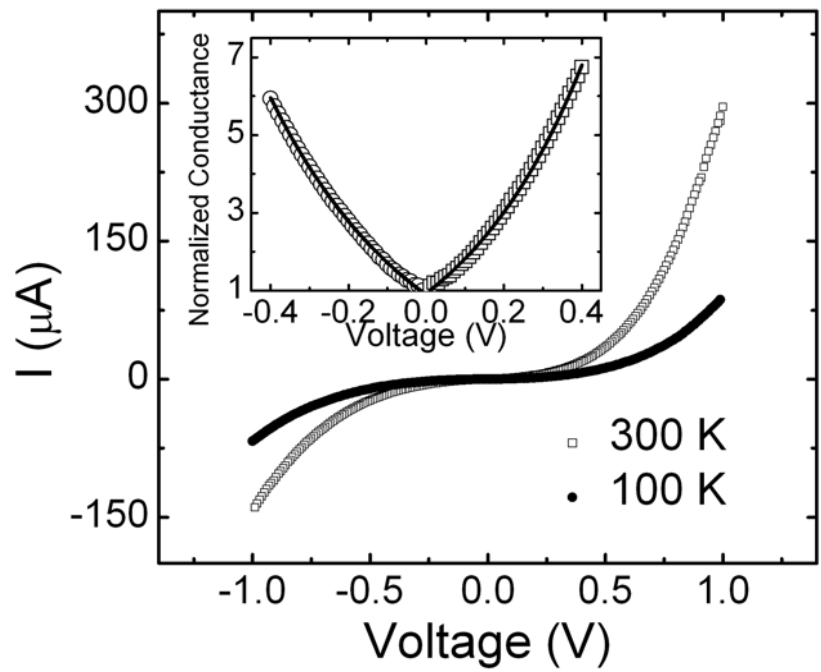

*Figure 2*

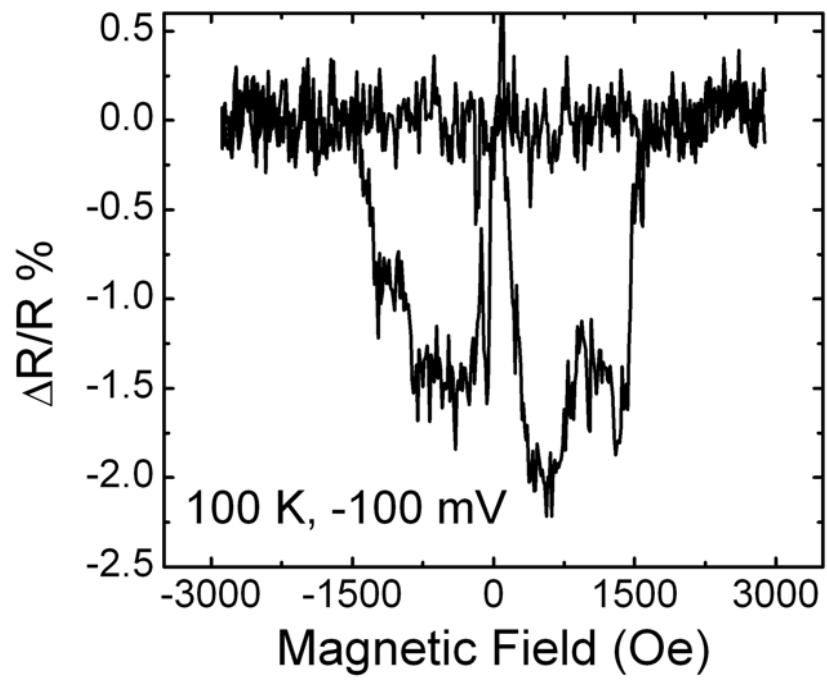

*Figure 3*

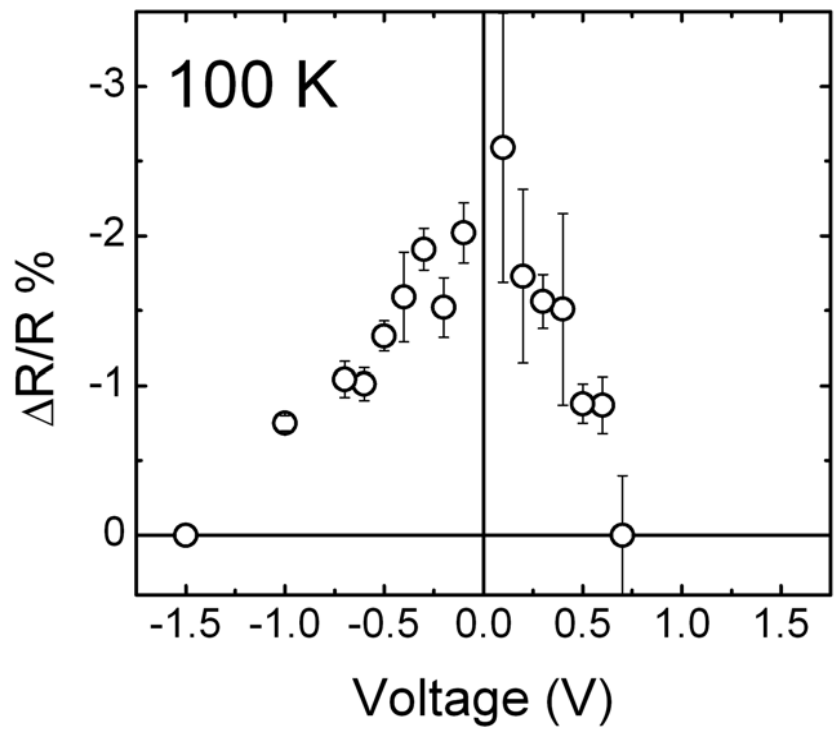

100 K

Figure 4

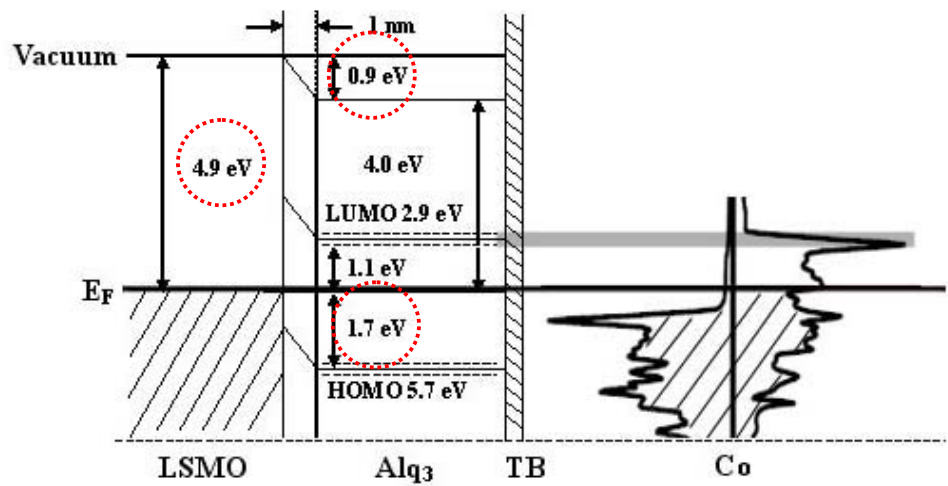